\documentclass[twocolumn,prX]{revtex4}

\usepackage{graphicx}% Include figure files
\usepackage{color}
\usepackage{dcolumn}% Align table columns on decimal point
\usepackage{bm}% bold math

\begin{document}

\title{Doping dependence of the Raman peaks intensity of graphene near the Dirac point}
\author{C. Casiraghi}
\date{\today}

\affiliation{Physics Department, Free University, Arnimallee 14, 14195 Berlin, Germany}

\begin{abstract}
%Raman spectroscopy is a powerful tool for investigating the properties of graphene. In particular, Raman spectroscopy is able to probe changes in the doping level of graphene. The shift of the Fermi energy is usually obtained by applying a gate voltage to graphene in back or top gating configuration. The Raman peaks change in position, width and intensity depending on the gate voltage. The same variations have been observed in the Raman spectra of pristine graphene, deposited on silicon substrate covered with 300 nm of silicon oxide ($Si/SiO_x$) and they have been attributed to un-intentional doping, produced by charged impurities.

Here we use pristine graphene samples in order to analyze how the Raman peaks intensity, measured at 2.4 eV and 1.96 eV excitation energy, changes with the amount of doping. The use of pristine graphene allows investigating the intensity dependence close to the Dirac point. We show that the G peak intensity is independent on the doping, while the 2D peak intensity strongly decreases for increasing doping. Analyzing this dependence in the framework of a fully resonant process, we found that the total electron-phonon scattering rate is $\sim$ 40 meV at 2.4 eV.
\end{abstract}

\maketitle

Graphene continues to attract interest because of its unique electronic properties \cite{Nov438(2005),Zhang438(2005),MorozovNov(2007)}, which make it a potential material for future
nanoelectronics \cite{chen,lemme}.
Graphene layers can be readily identified by elastic and inelastic light scattering, such as Raman \cite{acfPRL} and Rayleigh \cite{ray} spectroscopies. Raman spectroscopy is able to identify graphene and also to provide several information such as the amount of disorder \cite{ccPSS}, doping \cite{ccAPL,pisana,das} and the atomic arrangements at the edges \cite{ccnano,basko_edge}. Beside these practical applications, Raman spectroscopy in graphitic systems is extremely interesting because it involves resonant conditions \cite{steffi,baskoRS}, defect-induced processes \cite{steffi} and Kohn Anomalies (KA), which strongly affects the $\Gamma$-$E_{2g}$ and $K-A'_{1}$ modes \cite{ka}.

All carbons show common features in their Raman spectra in the 800-2000 cm$^{-1}$ region, the
so-called G and D peaks, which lie at around 1580 and 1360 cm$^{-1}$ respectively \cite{acfRS}. The D peak is due to the breathing modes of sp$^2$ rings and requires a defect for its activation \cite{tk}. The G peak corresponds to the $E_{2g}$ phonon at the Brillouin zone center \cite{acfRS}. The Raman spectra of graphite and graphene also show second-order scattering \cite{acfPRL}. The D peak and its second order peak (2D) are activated by resonance processes \cite{steffi,baskoRS}.

%The activation process for the D and 2D peaks is an inter-valley process and is as follows: i) a laser induced excitation of an electron/hole pair; ii) electron-phonon scattering with an exchanged momentum \textbf{q}$\sim$\textbf{K}; iii) defect scattering for the D peak; electron-phonon scattering with an exchanged momentum \textbf{-q} for the 2D peak; iv) electron/hole recombination.

%A resonant intra-valley process is possible too: this gives rise to the so-called D'peak, which can be seen around 1620 cm$^{-1}$ in defected graphite \cite{lespade}. The 2D' peak is the second order of the D' peak.

%Since 2D and 2D' peaks originate from a Raman scattering process where momentum conservation is obtained by the participation of two phonon with opposite wavevector (\textbf{q} and -\textbf{q}), they do not require the presence of defects to be activated.

Raman Spectroscopy can easily monitor doping in graphene, as reported in refs. \cite{pisana,das,kim}. Here, the electron or hole concentration was directly controlled by applying a gate voltage, in top or back-gate configuration, which produces a shift of the Fermi energy ($E_F$) from the Dirac point. The Raman spectrum shows the following variations with doping:

i) the G peak position increases for increasing $|E_F|$ and saturates for high doping \cite{pisana,das,kim}. This is due to the nonadiabatic removal of the Kohn anomaly at $\Gamma$\cite{naKA};

ii) the G peak Full Width at Half Maximum decreases for increasing $|E_F|$, and saturates when the electron-hole gap becomes higher than the phonon energy \cite{pisana,naKA}. This is due to the blockage of the phonon decay into electron-hole pairs due to the Pauli exclusion principle \cite{naKA}.

iii) the 2D peak position increases for p-doping, while it decreases for n-doping, for increasing $|E_F|$\cite{das}. The 2D and G peaks show a different doping dependence because the nonadiabatic effects are expected to be negligible on the 2D phonons, when measuring with visible energy \cite{das}.

iv) the ratio between the peaks intensity, I(2D)/I(G), decreases for increasing $|E_F|$ \cite{das}. In the framework of the fully Raman resonant process for the second order 2D peak, the intensity dependence on the doping is due to the electron-electron scattering contribution, which increases for increasing charge concentration \cite{acfINT}.

The same variations have been observed in the Raman spectrum of several pristine graphene samples, deposited on $Si/SiO_x$ substrate and they have been attributed to doping by charged impurities \cite{ccAPL}. Pristine graphene samples can have charge concentration up to $10^{-13}$ $cm^{-2}$, and they are usually p-doped \cite{ccAPL,ccPSS}. Unintentional doping in graphene deposited on $Si/SiO_x$ was first observed in gating experiments: the G peak position obtained for $E_F=0$ did not correspond to the G peak position of an undoped graphene ($\sim$ 1580 cm$^{-1}$, see Fig. 4 in ref. \cite{kim}). Thus, it is not possible to reach the Dirac point by gating due to local charge density variations \cite{kim}. Recent works \cite{susp,singa_susp,afmcharge} have finally confirmed the presence of charged impurities in graphene. In particular, Raman and transport measurements performed on suspended graphene have shown that by eliminating the substrate it is possible to produce graphene samples, which are essentially undoped, have little disorder and the highest mobility \cite{susp,bolotin}.

% \begin{figure}
%\centerline{\includegraphics [width=80mm]{doping1.eps}}
%\caption{First (a) and second (b) Raman spectrum of a pristine graphene deposited on a $Si/SiO_x$ substrate.}
%\end{figure}

Here we analyze the absolute intensity of the G and 2D peak, I(G) and I(2D) respectively, of several pristine graphene samples with different amount of doping. The use of pristine graphene samples presents the following advantages: i) low doping level can be investigated, in contrast to gated-graphene; ii) graphene is not in a device-configuration, so contacts and lithography processes do not affect the properties of graphene \cite{ccAPL}; iii) the sample is stable and it cannot be damaged by electrostatic charge or high voltage. This is particular important when measuring the absolute Raman intensity, since these measurements can be drastically altered as a result of surface conditions and strongly depend on the experimental set-up \cite{cross}. We will show that I(G) is insensitive to the doping, in contrast to I(2D), which strongly depends on the Fermi energy, as expected in the framework of the fully resonant activation process for the 2D peak \cite{baskoRS}.

We study several graphene samples, produced by micro-mechanical cleavage of bulk
graphite and deposited on $Si/SiO_x$. Only single layers flakes that completely fills the laser focus ($\sim$ 1 $\mu m^2$) were selected in order to avoid the graphene edges, since they can strongly affect the intensity \cite{ccnano}. Unpolarized Raman spectra are measured at 633 nm and 514 nm by using a Renishaw micro-Raman 1000 spectrometer. The Raman spectra are collected with a 100X objective and the spectral resolution is $\sim$ 3 cm$^{-1}$. The power on the samples is always kept well below 2 mW. The spectra have been fitted by using a Lorentzian spectral shape for both G and 2D peaks and the intensity is calculated as integrated area. The Raman spectra of the samples do not show any D peak, indicating a high crystallinity. In the case of pristine graphene samples, the amount of doping is not directly accessible through a gate voltage. Thus, we used the dependence of the G peak position on the doping in order to derive the Fermi energy for every sample. We used the relations between G peak positions and $E_F$ calculated for an ideal graphene at 300K in ref. \cite{naKA}.

%Figure 1 shows a typical Raman spectrum of graphene deposited on $Si/SiO_x$. The ratio I(2D)/I(G) is 6.5 and I(2D)/I(2D') is 34. From the G peak position, we derived a shift in the Fermi energy of $\sim$ 0.1 eV.

Figure 1 correlates I(G) and I(2D) with $E_F$. This figure clearly shows that I(G) is insensitive to changes in $E_F$, in agreement with the predictions in ref. \cite{basko_newjphys}. In contrast, I(2D) strongly decreases for increasing doping. This is in agreement with the Raman theory, under the assumption of a fully resonant process and for $E_F<< 1$ eV \cite{acfINT}. Assuming two main scattering mechanisms, i.e. the emission of phonons ($\gamma_{ep}$) and the electron-electron collision ($\gamma_{ee}$), I(2D) can be written as \cite{acfINT}:

\begin{equation}
    I(2D)= C (\gamma_K / \gamma)^2
    \end{equation}

where C is a constant and $2\gamma$ is the total scattering rate, so $\gamma=\gamma_{ep}+\gamma_{ee}$ and $\gamma_{ep}=\gamma_K + \gamma_\Gamma$. %A similar expression is valid for I(2D') so that $I(2D)/I(2D')=2(\gamma_K/ \gamma_\Gamma$)^2$ \cite{acfINT}.
 Note that the scattering rate $\gamma_K $ and $\gamma_\Gamma$ depends on the incident energy \cite{acfINT}.

 \begin{figure}
\centerline{\includegraphics [width=80mm]{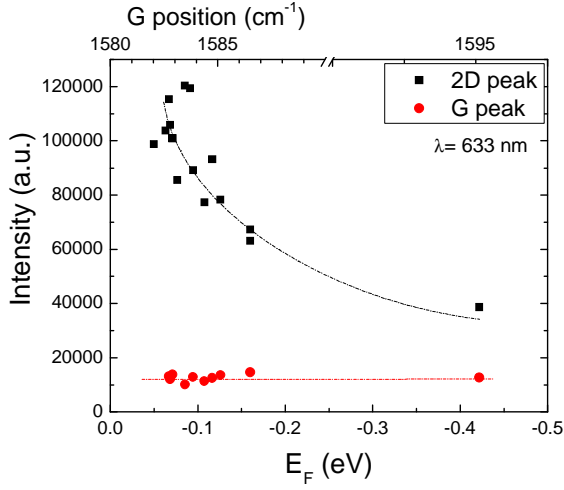}}
\caption{(Color online) G and 2D peak intensity as a function of the Fermi energy and G peak position, measured at 633 nm. The dotted lines are a guide for the eyes.}
\end{figure}

 \begin{figure}
\centerline{\includegraphics [width=80mm]{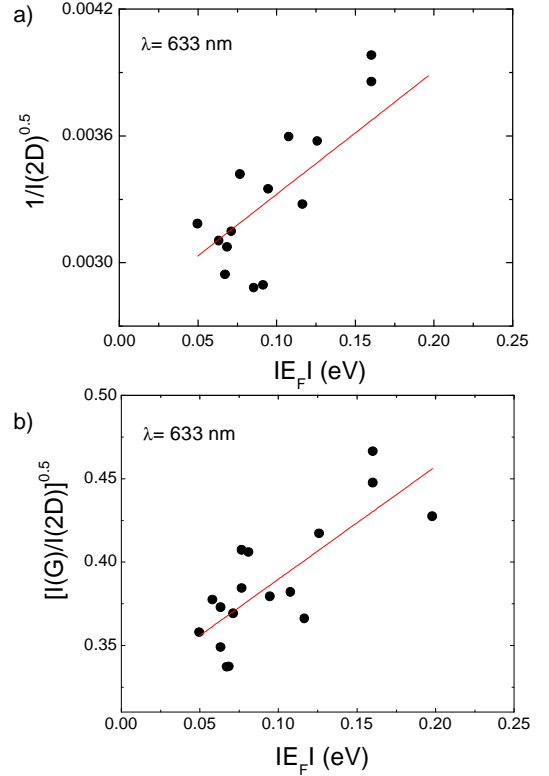}}
\caption{(Color online) Fit of the experimental dependence: (a) $1/ \sqrt{I(2D)}$ and (b) $1/ \sqrt{I(G)/I(2D)}$ as a function of the Fermi energy, measured at 633 nm. Both fits give $\gamma_{ep}\sim$ 30 meV.}
\end{figure}

 It has been shown that only $\gamma_{ee}$ depends on $E_F$, so eq. 1 can be written as:

\begin{equation}
    \sqrt{1/I(2D)}= \frac{1}{\gamma_K \sqrt{C} } (\gamma_{ep}+ f(\varepsilon)|E_F|)
    \end{equation}

where f is a function which depends on the dielectric environment \cite{acfINT}. In our case, the dielectric constant $\varepsilon$ for $SiO_x$ is 4.5, so $f\simeq 0.07$ \cite{acfINT}. Thus, eq. 2 shows that I(2D) decreases for increasing $E_F$, as observed in Figure 1.

Eq. 2 can be used to derive the electron-phonon scattering rate by measuring the variation of the absolute intensity of the 2D peak with doping. However, this requires to compare the absolute Raman intensity, i.e. to measure all the spectra under exactly the same experimental conditions. This is not always possible, in particular, during gating experiments. Since I(G) is insensitive to the doping (Fig. 1), I(G)/I(2D) can be used in order to derive $\gamma_{ep}$ since:

\begin{equation}
    \sqrt{I(G)/I(2D)}= C'(\gamma_{ep}+ 0.07|E_F|)
    \end{equation}

where C' is another constant. Eq. 3 has been used to find $\gamma_{ep}$ of gated-graphene \cite{acfINT}. There, values ranging from 18 to 65 meV, with an average of $\sim$ 33 meV, have been reported, depending on the data set \cite{acfINT}. In particular, for a hole-doped graphene a good consistency between different data has been found, resulting in $\gamma_{ep}$$ \sim$ 20 meV at 514 nm \cite{acfINT}. Here we use eqs 2 and 3 in order to find $\gamma_{ep}$ using pristine graphene samples. The present data, even though collected from many samples, are much less scattered than the data obtained by gating experiments, in particular close to the Dirac point, where eqs. 2 and 3 are well valid. Thus, our data enable determining  $\gamma_{ep}$ with a significantly smaller uncertainty.

 \begin{figure}
\centerline{\includegraphics [width=80mm]{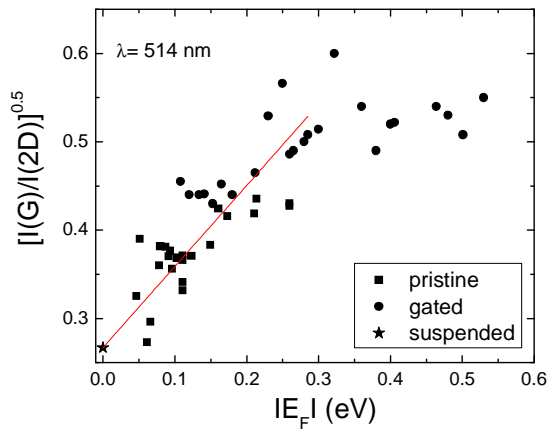}}
\caption{(Color online) Fit of the experimental dependence $1/ \sqrt{I(G)/I(2D)}$ as a function of the Fermi energy for pristine graphene, measured at 514 nm. The data for suspended graphene \cite{susp} and the back-gated graphene \cite{acfINT} have been included in the figure.}
\end{figure}

Figure 2 (a) shows the linear fit of the experimental dependence $\sqrt{1/I(2D)}$ on the doping (eq. 2), measured at 633 nm . We found: $\gamma_{ep}$= 33 meV. The linear fit of the experimental dependence $\sqrt{I(G)/I(2D)}$ on the doping (eq. 3), measured at 633 nm, is shown in Figure 2(b) and gives again $\gamma_{ep}$= 33 meV. Thus, for an intrinsic graphene, i.e. for $E_F$= 0, the fit in Fig. 2 (b) gives I(2D)/I(G) $\sim$ 10.

Figure 3 shows the linear fit of the experimental dependence $\sqrt{I(G)/I(2D)}$ on the doping (eq. 3), measured at 514 nm. Here, we included the data obtained for suspended graphene, by constraining the fit and we obtained $\gamma_{ep}$ = 21 meV. Furthermore, we included in Figure 3 the data obtained for graphene in back-gate configuration for $|E_F|> 0.1$ eV, as taken from ref. \cite{acfINT}. Figure 3 clearly shows that eq. 3 is valid up to $E_F$= 0.3-0.4 eV, as expected, since eqs. 2 and 3 are valid only for $E_F<<1 eV$.

In conclusion, we have shown that the G peak intensity is independent on the doping, while the 2D peak intensity is strongly sensitive to the dynamics of the photo-excited electron-hole pairs. By measuring the dependence of I(2D) on the Fermi energy close to the Dirac point, we have found that the total electron-phonon scattering rate (2$\gamma_{ep}$) of graphene is $\sim$ 40 meV at 2.4 eV, in good agreement with the hole-doping side of the top and back-gated graphene experiments \cite{acfINT}.

The author acknowledges D. M. Basko for inspiring discussions and critical reading of the manuscript. This work was supported by the Alexander von Humboldt Foundation in the framework of the Sofja Kovalevskaja Award, endowed by the Federal Ministry of Education and Research.

\end{document}